\documentclass[aip, apl, reprint, groupedaddress]{revtex4-1}
\usepackage{graphicx}
\usepackage{siunitx}
\usepackage{physics}

\begin{document}
\title{Theory of resonant tunneling in bilayer-graphene/hexagonal-boron-nitride heterostructures}
\author{Sergio C. de la Barrera}
\author{Randall M. Feenstra}
\affiliation{Department of Physics, Carnegie Mellon University, Pittsburgh, Pennsylvania, 15213, USA}

\begin{abstract}
A theory is developed for calculating vertical tunneling current between two sheets of bilayer graphene separated by a thin, insulating layer of hexagonal boron nitride, neglecting many-body effects.
Results are presented using physical parameters that enable comparison of the theory with recently reported experimental results.
Observed resonant tunneling and negative differential resistance in the current--voltage characteristics are explained in terms of the electrostatically-induced band gap, gate voltage modulation, density of states near the band edge, and resonances with the upper sub-band.
These observations are compared to ones from similar heterostructures formed with monolayer graphene.
\end{abstract}
\maketitle

%---------------------------------INTRODUCTION---------------------------------%
In contrast to the well-known linear dispersion of monolayer graphene (MLG), charge carriers near the six corners of the Brillouin zone in an isolated graphene \emph{bilayer} are described by a quadratic energy dispersion.\cite{mccann2006landau, mccann2006asymmetry}
An even more intriguing distinction with MLG is that, under the influence of external fields, the band structure of bilayer graphene (BLG) near the charge neutrality point becomes quartic, changing from semi-metallic to semiconducting as a small band gap is induced.\cite{min2007abinitio, castroneto2007biased, oostinga2007gate}
This tunability of the band gap can be exploited by introducing gates, doping, and interactions with substrate materials in electronic devices based on BLG.\cite{ohta2006controlling, zhang2009direct, lee2014chemical}
In this paper, we consider these effects and others in a 2D to 2D resonant tunneling device composed of two sheets of BLG separated by a thin, insulating layer of hexagonal boron nitride (h-BN).
In a vertical configuration with an interlayer bias, tunneling between two-dimensional electron gases is constrained by simultaneous energy and momentum conservation, leading to resonances in the current--voltage ($I$--$V$) characteristic and thus regions of negative differential resistance (NDR).\cite{fallahazad2015gate}
Such devices were originally proposed for conventional 2D quantum wells,\cite{simmons1998planar} but the theory was recently treated for MLG,\cite{feenstra2012single, zhao2012symfet, kumar2012modeling, sensale-rodriguez2013graphene, ryzhii2013dynamic, britnell2013resonant, delabarrera2014theory, brey2014coherent} and NDR was observed experimentally in high-quality devices shortly thereafter.\cite{britnell2013resonant, mishchenko2014twist, fallahazad2015gate}
The theory discussed in the present work is particularly relevant to the recent observations of Fallahazad et al.\cite{fallahazad2015gate}

\begin{figure}
\includegraphics{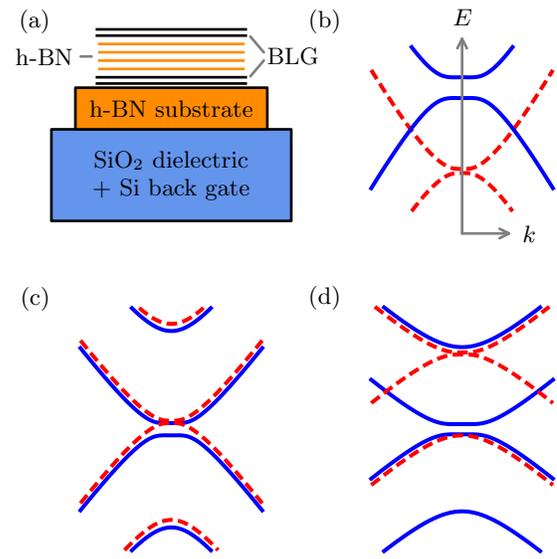}
\caption{
(a) Device structure, with double black lines indicating each graphene bilayer (BLG) and the group of orange lines representing 4 to 6 layers of h-BN.
(b) Alignment of electronic bands at an off-resonant interlayer bias voltage; blue (solid) curves for one bilayer and red (dashed) for the other; (c) at the voltage which yields the primary tunneling resonance; (d) at a higher voltage which aligns the lower bands of one bilayer with the higher sub-bands of the other.
The largest contribution to tunneling current occurs near the states where the two bands intersect.
Bands represent energy as a function of in-plane crystal momentum near one of the six corners of the Brillouin zone.
\label{fig:overlap}}
\end{figure}

%----------------------------------MECHANISM-----------------------------------%
For a given interlayer voltage and for bilayers that are in crytallographic alignment, the electronic bands of the top and bottom bilayer of graphene will overlap for particular sets of states with equal energy and crystal momentum (Fig.~\ref{fig:overlap}).
Away from the resonance voltage, only the states near the intersecting ring(s) can contribute to the tunneling current (Fig.~\ref{fig:overlap}b).
However, for one particular voltage, the electrostatic potential between the bands will be zero, allowing all states between the two Fermi levels to tunnel simultaneously (Fig.~\ref{fig:overlap}c).
The shape and position of the resulting resonant peak in the $I$--$V$ characteristic depends on the quantity and sign of charge carriers in each bilayer, and therefore indirectly on external fields (gate voltages) and the electrostatic doping conditions.

\begin{figure}
\includegraphics{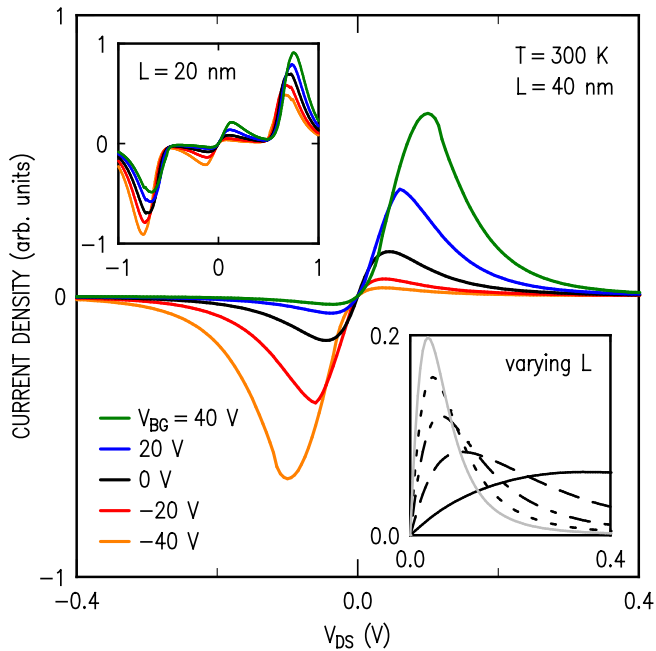}
\caption{
Calculated tunnel current density versus interlayer bias for undoped graphene bilayers for a range of gate voltages.
Upper inset: a similar computation for a larger voltage range highlighting secondary resonances from the higher sub-bands.
Lower inset: a closer view of the $V_\text{BG} = 0$ case, varying coherence length, a disorder parameter in the theory, from \SI{50}{nm} (solid, light) to \SI{10}{nm} (solid, dark).
\label{fig:high-voltage}}
\end{figure}

For example, in the absense of strong doping or substrate interactions, resonance can be observed for both both positive and negative bias voltages as the back-gate voltage ($V_\text{BG}$) is swept from one sign to the other (Fig.~\ref{fig:high-voltage}).
Recently, Fallahazad et al. have observed resonances with precisely this behavior in devices fabricated with exfoliated BLG/h-BN/BLG on a h-BN/SiO$_2$ substrate.\cite{fallahazad2015gate}
The width and amplitude of each resonant peak relative to the background (non-resonant) current are determined by the degree of coherence between tunneling wavefunctions, as is discussed in detail in Ref.~\onlinecite{delabarrera2014theory}.

In addition to the primary resonance, the higher sub-band of one bilayer can also come into alignment with the lower sub-band of the second bilayer causing a similar spike in the tunneling current.
Secondary resonances as well as an increase in background current from the upper bands entering the tunneling energy window can be observed at larger voltages as shown in the upper inset of Fig.~\ref{fig:high-voltage}.
Interactions with the upper sub-bands are distinct from monolayer devices, and may provide opportunities for multi-state logic devices.

\begin{figure}
\includegraphics{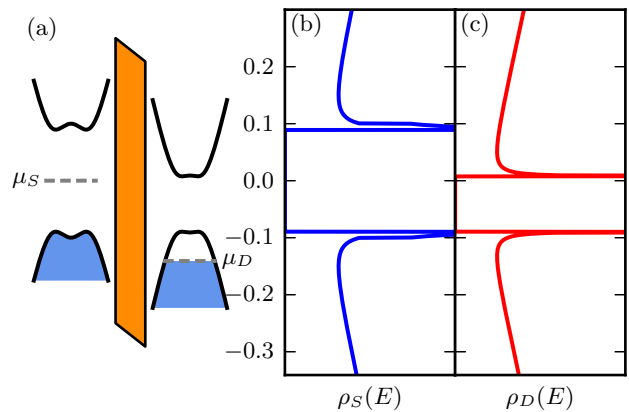}
\caption{
(a) Electronic bands in the source (left) and drain (right) electrodes with the tunneling barrier (band gap of boron nitride) in between at a small positive bias.
Dashed lines indicate the Fermi levels in each bilayer, $\mu_i = -eV_i$; not to scale.
Density of states corresponding to (b) the source electrode and (c) the drain electrode in the same bias condition as panel (a); energy axes in units of \si{eV}.
The alignment of the divergences in the density of states near the valence band edge of each bilayer produces a large overlap of states and thus a spike in tunneling current.
\label{fig:bandgap}}
\end{figure}

\begin{figure}
\includegraphics{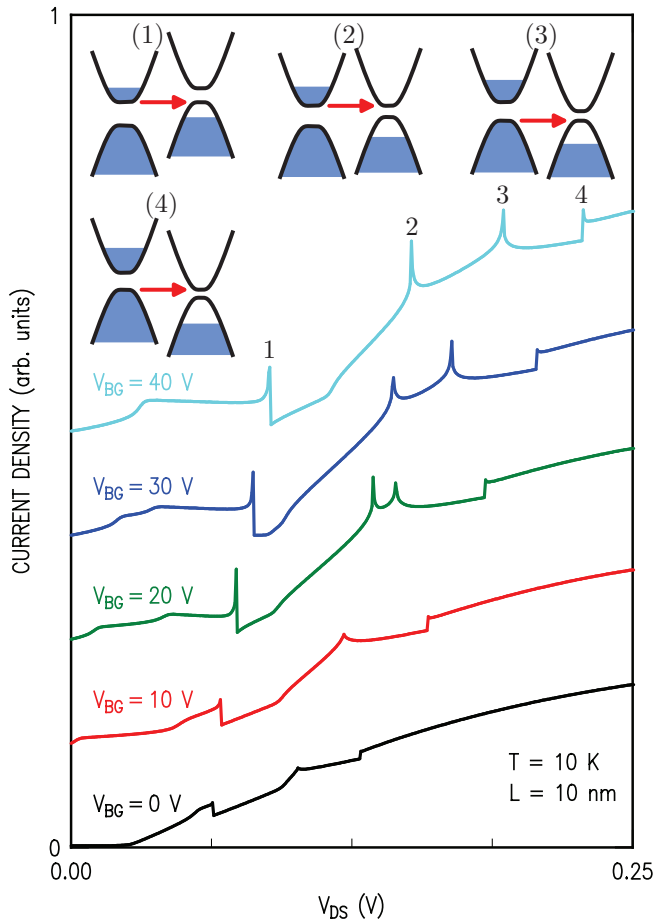}
\caption{
Low-voltage tunneling current for a device with a small amount of doping on the top (drain) bilayer at \SI{10}{K} showing a number of small features due to the alignment of various band-edges, as explained in Fig.~\ref{fig:bandgap}.
$I$--$V$ curves are shifted vertically for clarity.
Numbered insets show the band alignment for each of the four labeled points along the $V_\text{BG} = \SI{40}{V}$ curve.
Arrows indicate electron current that produces the sharp feature in each case.
\label{fig:low-voltage}}
\end{figure}

At a smaller voltage scale, and especially at lower temperatures, it is possible to observe additional features due to the tunable band gap in BLG.
The presence of a transverse electric field across a graphene bilayer induces a potential difference between the two individual layers of graphene.
This broken layer symmetry causes a small band gap to open up around the charge neutrality point which increases with the magnitude of the potential difference across the bilayer.
In the tunneling device modeled here, the interlayer and gate voltages modulate the separate potential difference across each individual bilayer in a coupled system.\cite{lee2014chemical}
As a result, the band gaps in both bilayers vary with voltage (typically at different rates), which affects the overall tunneling current. %% WEAK SENTENCE
For non-zero band gap, the precise form of the energy dispersion is quartic near the gap, as in Figs.~\ref{fig:overlap}~and~\ref{fig:bandgap}(a).
Moreover, the location of the band gap is a ring of states concentric with the $K$-point. %% INTRODUCE K-POINT EARLIER
This arrangement of states causes divergences in the DOS at the conduction and valence band edges, which can yield additional spikes in the tunneling current for certain electrostatic arrangements.
Whereas the primary feature in the tunneling current occurs when the electrostatic potentials in the source $\phi_S$ and drain $\phi_D$ electrodes are aligned, $\Delta\phi = \phi_S - \phi_D = 0$, other features due to overlap of the large DOS near the band edges can occur when one of the four conditions $\Delta\phi \pm E_{g,S}/2 \pm E_{g,D}/2 = 0$ is satisfied (where $E_{g,i}$ is the band gap in each bilayer), as in Figs.~\ref{fig:bandgap}~and~\ref{fig:low-voltage}.
These features in the $I$--$V$ characteristic are distinct from those caused directly by momentum-conserving effects with complete band alignment (as in Fig.~\ref{fig:overlap}) and are less sensitive to the relaxation of momentum conservation (decoherence), but may be observed in tandem with the latter.
In MLG there are no equivalent band edges, and thus these additional sharp features are absent in monolayer vertical tunneling devices.

%----------------------------------FORMALISM-----------------------------------%
We use a tight-binding model for the dispersion of BLG with nearest-neighbor hopping energy $\gamma_0 \approx \SI{3.1}{eV}$, interlayer hopping energy $\gamma_1 \approx \SI{0.4}{eV}$, and interlayer potential asymmetry $U$.\cite{mccann2013electronic}
Higher order considerations such as the trigonal warping of the bands (azimuthal asymmetry) were found to have a negligible impact on the tunneling and thus were excluded.
The occupation of levels and band gap in each electrode varies with the set of applied voltages, and thus the electrostatic potentials are required to calculate the tunneling current.
These potentials are determined by first solving a matrix equation $q_i = C_{ij}V_j$, treating each monolayer of graphene separately, to obtain the transverse fields across each bilayer.
We then use those fields to solve a second matrix equation treating each bilayer with the local DOS for each layer within the bilayers.
This method can accomodate both top and bottom gates, though we chose to focus on matching with devices with only one gate in the present work.
Net charges are calculated using full Fermi integrals $q_i = e(n_i - p_i) - e(n_{0,i} - p_{0,i})$, $n_i = e\int\dd{E}\rho(E)f(E)$ to account for quantum capacitance and thermal occupation, with environmental doping densities $n_{0,i}$.
We calculate the tunneling current by summing over the transition rates between all states in the source and drain bilayers,
\begin{equation}
I = g_s g_v \frac{2\pi e}{\hbar} \sum_{\alpha,\beta} \abs{M_{\alpha\beta}}^2 \bqty{f_S(E_\alpha) - f_D(E_\beta)} \delta(E_\alpha - E_\beta),
\label{eq:current}
\end{equation}
with spin and valley degeneracies $g_s$, $g_v$ and state labels $\alpha$ and $\beta$ in the source and drain bilayers.\cite{feenstra2012single}
The overlap integrals between states in the source and drain are contained in the matrix element
\begin{equation}
M_{\alpha\beta} = \frac{\hbar^2}{2m} \int \dd{S} \pqty{\Psi_\alpha^* \dv{\Psi_\beta}{z} - \Psi_\beta \dv{\Psi_\alpha^*}{z}},
\label{eq:matrix-el}
\end{equation}
which is evaluated in a similar way as for MLG in Refs. \onlinecite{feenstra2012single} and \onlinecite{delabarrera2014theory}.
We calculate the surface integral in Eq. \ref{eq:matrix-el} over a region defined by the length scale of wavefunction coherence in the device, a parameter we call the characteristic coherence length, $L$.
This is a disorder parameter which defines the degree of momentum conservation and thus controls the width and amplitude of resonant features in the $I$--$V$ characteristic.
The momentum (wavevector) conservation, chiral (angular) terms, and crystallographic misorientation are encapsulated in the matrix elements, while energy conservation is contained in the $\delta$-function that appears in Eq.~\ref{eq:current}.\cite{feenstra2012single, delabarrera2014theory}

In contrast to the theory for MLG devices, for BLG this $\delta$-function must be evaluated using the quartic dispersion relation in order to capture band-gap and higher sub-band effects.
Converting the sums over states in Eq. \ref{eq:current} to integrals over $k$, we can evaluate the $\delta$-function by changing variables from $E$ to $k$,
\begin{equation}
\delta[E(\vb{k}_S) - E(\vb{k}_D) + \Delta\phi] \rightarrow \sum_i \frac{\delta(k - k_i)}{\abs{f'(k_i)}}
\end{equation}
for all combinations of bands between each bilayer, where $f$ is equal to the original argument of the $\delta$-function, $k_i$ are the zeros of this argument, and $f'$ is the derivative with respect to $k$.
This procedure allows us to remove one $k$-integration and proceed to calculating the current.
A small amount of broadening is introduced to handle the singularities that arise near the band edges (an imaginary term $i\epsilon$ is added to the $\abs{f'}$ denominator, with epsilon typically equal to $10^{-2}\hbar v_F$).

%------------------------------COMPARISON TO EXP-------------------------------%
Comparing our theory with the experimental results of Fallahazad et al.\cite{fallahazad2015gate}, we find for the undoped device at room temperature (Fig.~\ref{fig:high-voltage}) very good agreement both in terms of the peak shapes and the gate-voltage dependence.
For the low-temperature results of Fig.~\ref{fig:low-voltage}, small peaks associated with DOS features become prominent, superimposed on a broad momentum-conserving background current.
We believe the situation found in experiment at low temperature is the same, showing a similar sharp peak superimposed on a smooth background current.\cite{fallahazad2015gate}
The interpretation offered in Ref.~\onlinecite{fallahazad2015gate} associates the sharp peak itself with a momentum-conserving resonant effect, but no origin for the broad background is provided.
Alternatively, in our interpretation, both features can be well understood.
The data for the undoped device at room temperature (Fig.~\ref{fig:high-voltage}) can be similarly understood within the same framework.
Sharp DOS features are not seen for the latter, either in theory or experiment, since the higher temperature leads to a reduction in the amplitude of the sharp peaks (at elevated temperature the tunnel current includes contributions from nearby states that are thermally occupied, leading to a reduction in strength of the sharp peaks).
This distinction between DOS versus momentum-conserving effects, as provided by our theory, provides an expanded interpretation of the experimental results.\cite{fallahazad2015gate}

%---------------------------------CONCLUSIONS----------------------------------%
While resonant tunneling in MLG heterostructures is novel and intriguing, the additional sub-bands in BLG as well as its unusual behavior in the presence of transverse fields provides many additional channels for interesting tunneling phenomena.
Although the results presented here were calculated with zero angular misorientation (perfect crystallographic alignment) between the two bilayers of graphene, the theory readily computes current for non-zero misorientation, as discussed and observed in prior work for MLG.\cite{feenstra2012single, delabarrera2014theory, mishchenko2014twist}
Concerning possible misorientation within the graphene bilayers themselves, this is known experimentally not to occur for the devices of Fallahazad et al.\cite{fallahazad2015gate}
An additional source of misorientation in the device would be that between the graphene bilayers and the h-BN layers of the tunnel barrier.
We have not investigated this effect in detail, although referring to prior work for twisted BLG,\cite{perebeinos2012phonon, masumhabib2013coherent} it appears that such an effect would give rise to a reduced transmission current through the entire heterostructure.
Indeed, for the case of tunneling between MLG layers separated by h-BN, computed tunnel currents agree in detail with experiment, except that the theory is a factor of $10^3$ to $10^4$ too large.\cite{delabarrera2014theory}
We find a similar discrepancy in absolute magnitude of the current for the present situation of BLG/h-BN devices, and we consider it likely that the reduced conductance of the BLG/h-BN interface is the source of this discrepancy.

For BLG devices, we find that DOS effects are largely unaffected by small amounts of angular misorientation between the bilayers, whereas momentum-conserving resonant peaks are shifted due to the change in conditions required for band intersection, as in monolayer devices.
We note that the electronic properties of the BLG can be expected to be influenced by the neighboring h-BN, in analogy with the MLG case.\cite{chen2014observation}
Such effects are typically on the 1 to \SI{10}{meV} scale; they will be important for a very detailed comparison between experiment and theory, but in terms of the overall distinction that we make here between DOS and momentum-conserving effects these effects can be neglected.
Similarly, we neglect many-body modifications to the BLG band structure (including many-body effects between the two graphene bilayers, since they are separated in the experiments\cite{fallahazad2015gate} by 4 to 6 monolayers of h-BN).
The effects of external in-plane magnetic fields have been explored for similar monolayer and monolayer/bilayer devices,\cite{pratley2014valley, brey2014coherent, mishchenko2014twist} but are not considered here for brevity.
Finally, inelastic effects may play a role in some devices, particularly at room temperature, however, we have focused here on elastic interactions, which play a large role in the relaxation of momentum conservation and subsequently the strength of resonant behavior compared to background current.

%-------------------------------ACKNOWLEDGEMENTS-------------------------------%
\section*{Acknowledgements}
We would like to thank Emanuel Tutuc for useful discussions.
This work was supported in part by the Center for Low Energy Systems Technology (LEAST) one of six centers of STARnet, a Semiconductor Research Corporation program supported by MARCO and DARPA.

%----------------------------------REFERENCES----------------------------------%
\section*{References}
\bibliography{refs}
\end{document}